# Growth of $Sr_{1-x}Ca_xRuO_3$ thin films by metalorganic aerosol deposition

**M Schneider, V Moshnyaga, P Gegenwart**

Georg-August-Universität, Friedrich-Hund-Platz 1,Göttingen 37073, D

E-mail: mschnei@uni-goettingen.de

**Abstract**. We report the growth of thin films of $Sr_{1-x}Ca_xRuO_3$ on $SrTiO_3$ and MgO substrates by metalorganic aerosol deposition. The structure and microstructure is characterized by X-ray diffraction and room-temperature scanning tunnelling microscopy (STM), respectively. STM indicates in-plane epitaxy and a small surface roughness for films on $SrTiO_3$. The high-quality of the films is supported by large residual resistivity ratios up to 29.

## 1. Introduction

$SrRuO_3$, with distorted $GdFeO_3$-type perovskite structure [1], is an itinerant ferromagnet [2]. This material is of great current interest, because the ferromagnetic ordering temperature, $T_C$ ~160K can be suppressed in $Sr_{1-x}Ca_xRuO_3$ and a quantum phase transition (QPT) emerges near x=0.7 [3]. Due to the smaller ionic radius of Ca compared to Sr-ions [4] the lattice distortion increases, which, however, is not of Jahn-Teller type; there exist only $RuO_6$ octahedra with constant bond lengths, tilted along (100)-axis [1]. The nature of the QPT is controversially discussed [3,5-7] and high-quality samples for low-temperature experiments are badly needed. Below, we report the successful growth and structural characterization of $Sr_{1-x}Ca_xRuO_3$ high-quality epitaxial thin films, whose low-temperature physical properties are discussed in [8].

## 2. Metalorganic aerosol deposition and preparation of films

$Sr_{1-x}Ca_xRuO_3$ thin films with x=0-1 have been prepared by metalorganic aerosol deposition (MAD), which was shown to be advantageous for the growth of multicomponent oxide films of different functionality [9]. Mixtures of acetylacetonates with the desired metal ions were dissolved in an organic solvent (e.g. dimethylformamide) and the solution was sprayed onto a heated substrate by means of a pneumatic nozzle with a carrier gas (e.g. dried air). In the close proximity to the heated substrate the precursor gas phase undergoes a heterogeneous pyrolysis reaction. The metal-acetylacetonates decompose at $T_{sub}$~250-300°C with subsequent oxidation of metal ions on the substrate surface with formation of an oxide film of desired composition.

For preparation of $Sr_{1-x}Ca_xRuO_3$ films, commercial acetylacetonates (acac) of $Sr^{2+}$, $Ca^{2+}$ and $Ru^{3+}$ with chemical purity of 97% ($Ru(acac)_3$, $Sr(acac)_2$) and 99,95% ($Ca(acac)_2$) were used. Precursor solutions with stoichiometric (Ca/Ru=1) and non-stoichiometric (Sr/Ru=1.4) compositions have been prepared in concentrations n=0.05-0.1M. Commercial one-side polished $SrTiO_3$ (STO) substrates of (100) orientation and 0.5mm×5mm×10mm size have been used. For comparison MgO (100) substrates were also employed. A substrate temperature of $T_{sub}$~900°C and aerosol flux rate of v=1.8ml/min were main typical deposition conditions for the preparation of epitaxial $Sr_{1-x}Ca_xRuO_3$ films of thickness d=40-50 nm. Note a high resulting deposition rate, r=2-3 nm/s. After preparation, the films were cooled down to room temperature in 30-45min. All samples look shiny and smooth, and exhibit no defects as visualized by optical microscopy.

For the films of SrRuO$_3$ on STO, a high residual resistivity ratio (RRR) up to 29 [8] and smooth surface morphology were obtained with a non stöchiometric precursor ratio of Sr/Ru=1.4. A possible reason for this ratio could be the different solubility of Sr- and Ru-precursors in dymethilformaimde. In general, as was shown for La-Ca-Mn-O thin films [9], the use of solutions with low molarities (0.01-0.02 M) can be advantageous. However, a low molarity was not suitable for SrRuO$_3$ films on STO. It resulted in an island growth with poor connections between crystalline grains and very low values of RRR~2 for such films. Possibly the solvent vapor chemically reacts with the growing film material, etching parts of the film during the deposition. In the case of low concentrations of precursors the removal rate seems to exceed the growth rate so that the substrate cannot be covered completely. On the other hand, solutions with n=0.1M are not usable because of oversaturation. We found, that an appropriate value of n=0.05M leads to continuous coverage of the substrate. The optimal deposition temperatures were found in the range T$_{sub}$=900-940°C. Lower temperatures result in the formation of particles on the films surface, ascertainable in scanning tunneling microscopy (STM). The highest tested temperature was close to the melting point of conductive silver paste, used to obtain a good thermal coupling between the substrate and the heater. To prepare high-quality CaRuO$_3$ thin films on STO with RRR=16 [8] similar substrate temperatures, precursor solution molarities as well as deposition rates were used. Here the Ca/Ru ratio was found to be extremely important. Any deviations from the stoichiometric ratio Ca/Ru=1 cause a drastic loss in the film quality, i.e. the RRR was found to be significantly reduced.

### 3. X-ray diffraction analysis

**Table 1**. Lattice plane distance d$_{110}$ for SrRuO$_3$ and CaRuO$_3$ thin films grown on STO and MgO, compared to bulk data [10]. f$_{MgO}$ and f$_{STO}$ denote lattice mismatch compared to substrates.

| x | $d_{110}^{MgO}$ [Å] | $d_{110}^{STO}$ [Å] | $d_{110}^{Bulk}$ [Å] | f$_{MgO}$ [%] | f$_{STO}$ [%] |
|---|---|---|---|---|---|
| 0 | 3.927 | 3.945 | 3.929 | 6.79 | 0.74 |
| 1 | 3.840 | 3.819 | 3.850 | 8.85 | 2.08 |

XRD θ-2θ scans demonstrate epitaxial growth of Sr$_{1-x}$Ca$_x$RuO$_3$ thin films on STO with (110)-axis, indexed in an orthorhombic P$_{nma}$ structure, perpendicular to the substrate surface. No secondary phases were detected. In Fig.1a) an example of XRD pattern for a representative SrRuO$_3$/STO film is shown. The film orientation is related to an orthorhombic structure and STO orientation to a cubic structure. Not marked additional reflexes are due to the substrate (Cu-K$_\beta$ lines) and the sample holder (Al). XRD patterns for films with x>0 are very similar with the only difference that the film peaks are progressively shifted to higher values of 2θ, evidencing a decrease of the lattice constant with increasing x. The corresponding dependence of the lattice constant d$_{110}$ on the Ca-concentration, x, is presented in the inset of Fig. 1a) for the films grown on STO. These changes in the crystal structure originate from smaller ionic radius of Ca$^{2+}$ compared to Sr$^{2+}$. The d$_{110}$-values for 0.1≤ x ≤ 0.4 could not be calculated, because of the close proximity of the sample reflexes to those of the STO substrates.

SrRuO$_3$ on MgO also grows in (110)-orientation without secondary phases (see Fig. 1b)). The unlabeled reflexes originate from the substrate and sample holder, too. Pure single-phased Sr$_{1-x}$Ca$_x$RuO$_3$ films (x>0) could hardly be grown on MgO. Secondary phases, as well as an additional orientation appear. This behaviour is presumably caused by the different film-substrate lattice mismatch for STO and MgO substrates. In Table 1 the d$_{110}$ values for SrRuO$_3$ and CaRuO$_3$ on the two different substrates are listed in comparison with d$_{110}$ values for bulk materials. From the XRD θ-2θ patterns the lattice mismatch is derived which can provide information on the lattice strain in the film. Samples on MgO grow nearly unstrained because of the formation of misfit dislocations. Indeed the differences between $d_{110}^{MgO}$ and $d_{110}^{Bulk}$ are very small. In contrast, the films on STO substrate, which have a small lattice mismatch, seem to be fully strained, yielding a larger difference of $d_{110}^{STO}$ compared to the bulk.

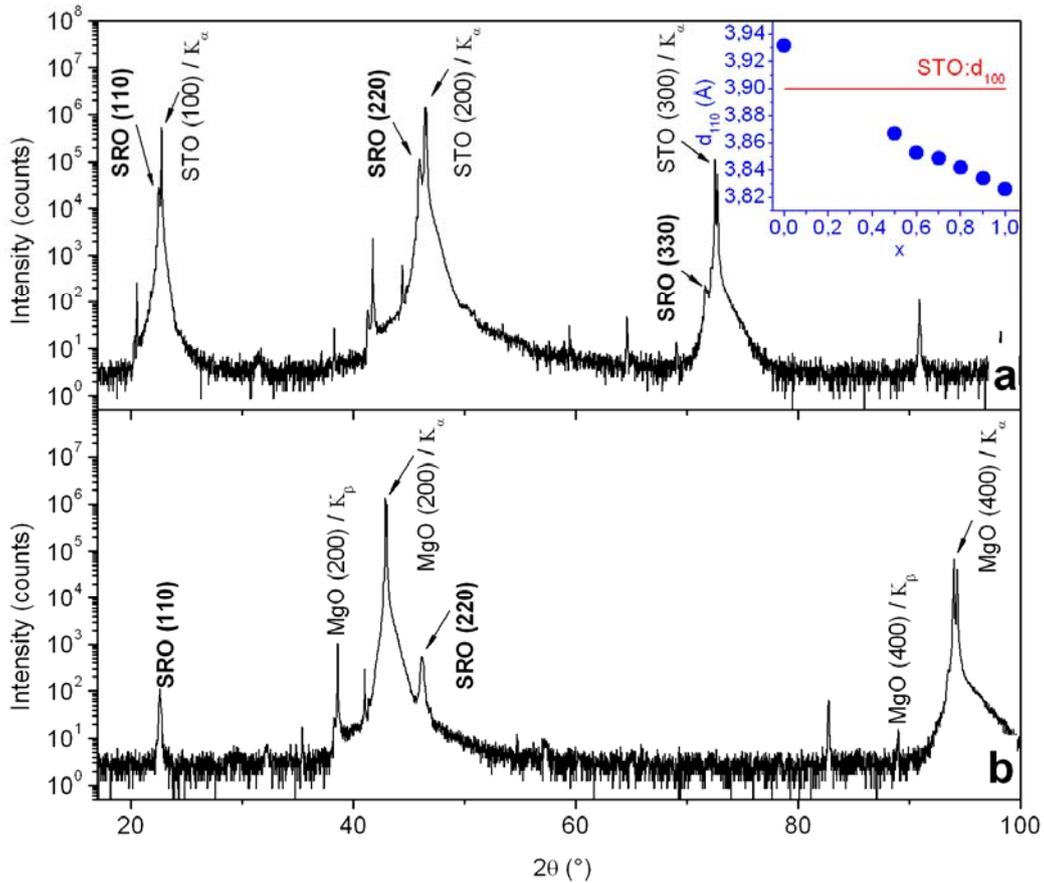

**Figure 1.** XRD θ-2θ spectra for SrRuO$_3$ on STO (a) and SrRuO$_3$ on MgO (b). The inset shows the x dependency of the lattice plane distance d$_{110}$. The red line indicates d$_{100}$ of STO.

## 4. Room temperature STM

Since in-situ electron scattering experiments (RHEED, LEED) are impossible for MAD which works at atmospheric pressure, we have studied the growth mode and surface morpholoogy using room temperature STM. In Figs. 2 a) and b) two STM morphology images of SrRuO$_3$ films grown on STO are shown. The morphology in Fig. 2 b) is typical for a layer-by-layer growth mode with atomically smooth terraces and steps of one atomic layer height. The calculated mean square roughness, RMS, in Fig. 2 b) is as small as 0.4 nm. It is interesting to note that layer-by-layer growth is possible within MAD but the small values of RMS are not necessary for high RRR values. Fig. 2 a) reveals rectangular blocks which seem to be well connected and possible free of defects as may be deduced from the high RRR=29 [8]. The crystalline blocks are apparently in-plane oriented, indicating an epitaxial growth. In-plane epitaxy shown by oriented rectangular blocks also appears for Sr$_{1-x}$Ca$_x$RuO$_3$ thin films for all Ca concentrations. In Figs. 2 c) and d) the images of CaRuO$_3$ and Sr$_{0.8}$Ca$_{0.2}$RuO$_3$ films are shown, respectively. One cannot definitely conclude layer-by-layer growth in the case of x>0. This may be caused by the lattice mismatch and/or the lack of a growth parameter optimization. Thin films grown on MgO are not of comparable quality. In-plane epitaxy cannot be concluded from STM measurements. Islands with heights of ~60 nm and a very high RMS of ~10 nm are observed. The reason is very probably the lattice mismatch of 7-9%, which results in 3d island growth.

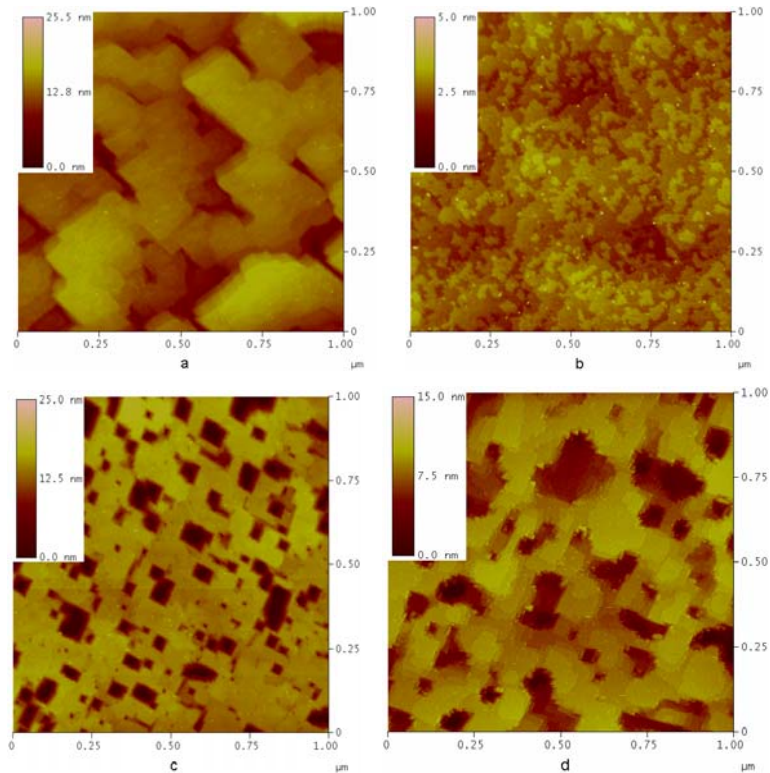

**Figure 2.** Examples for thin film surfaces measured with STM. $SrRuO_3$ (a) and (b), $CaRuO_3$ (c) and $Sr_{0.8}Ca_{0.2}RuO_3$ (d) all grown on STO.

## 5. Summary
We have grown $Sr_{1-x}Ca_xRuO_3$ thin films on STO substrates using MAD. The films are of high quality, i.e. they possess high residual resistivity ratio up to 29 and, for small x, where the lattice mismatch to STO is small a roughness as low as 0.3nm, comparable with the lattice constant. STM reveals an in-plane orientation of crystalline blocks indicating epitaxial growth. The $d_{110}$ lattice parameter decreases with increasing Ca-concentration x. The important deposition parameters are a high molarity of the precursor solution, n=0.05-0.1, a substrate temperature of $T_{sub}$~900-940°C, a high growth rate of ~2 nm/s, and precursor ratios Sr/Ru=1.4 and Ca/Ru=1, respectively.

**Acknowledgments** Work supported by the German Science Foundation DFG through SFB 602.